\documentclass[conference]{IEEEtran}
\IEEEoverridecommandlockouts

\usepackage{cite}
\usepackage{amsmath,amssymb,amsfonts}
\usepackage{algorithmic}
\usepackage{graphicx}
\usepackage{textcomp}
\usepackage{xcolor}
\usepackage{listings}

\def\BibTeX{{\rm B\kern-.05em{\sc i\kern-.025em b}\kern-.08em
    T\kern-.1667em\lower.7ex\hbox{E}\kern-.125emX}}
\begin{document}

\title{An Electronic Ising Machine\\}

\author{
\IEEEauthorblockN{Matt Bowring}
\IEEEauthorblockA{\textit{Dept. of Mechanical Engineering}}
\IEEEauthorblockA{\textit{Purdue University}}
\text{mbowring@purdue.edu}
\and
\IEEEauthorblockN{Ben Anderson}
\IEEEauthorblockA{\textit{Dept. of Electrical Engineering}}
\IEEEauthorblockA{\textit{Purdue University}}
\text{ande1448@purdue.edu}
\and
\IEEEauthorblockN{Ben Tiffany}
\IEEEauthorblockA{\textit{Dept. of Electrical Engineering}}
\IEEEauthorblockA{\textit{Purdue University}}
\text{btiffan@purdue.edu}
}

\maketitle

\begin{abstract}
We develop a custom printed circuit board (PCB) for a low-power and high-speed accelerator of NP-Hard graph problems. The architecture implements an annealing-based computing paradigm using a network of nonlinear electronic oscillators whose phase dynamics converge to stable configurations that encode solutions. We review the theoretical framework, and present our circuit design, simulations, and experimental results. We further highlight some key future research directions for the emerging developing of computing architectures based on energy minimization.

\end{abstract}


\section{Introduction}

Ising machines refer to a broad class of devices designed to solve the Ising model equation 

\begin{equation}
H = \mathbf{s}^T \mathbf{J} \mathbf{s}\label{ising}
\end{equation}

where $\mathbf{s} = (s_1, s_2, \dots, s_N)^T$ is the vector of $N$ binary spin variables $s \in \{-1, +1\}$, and $\mathbf{J} \in \mathbb{R}^{N \times N}$ is the symmetric coupling strength matrix. It was first studied in 1922 by Ernest Ising to represent the energy potential of interacting ferromagnets. Because of the exponential search space, finding spin configurations that minimize $\eqref{ising}$ is NP-Hard. This task can be equivalently used to represent many graph problems \cite{b1} taking the form of quadratic unconstrained binary optimization \cite{b2}. These problems quickly become intractable yet have deep applications in practical domains like scheduling logistics, finance, and machine learning to name a few.  

Recently, the search to scale compute “beyond Moore's Law” has focused research on coupled oscillator systems \cite{b3}. An Ising machine is the term coined to describe a device that minimizes the Ising model $\eqref{ising}$ by the convergence of some physical system to equilibrium through an annealing process \cite{b4}. The logical spin variables are represented analogously in a physical system using lasers, photonics, electronics, nanomagnets, qubits, or even fully digital programs. However, there remains fundamental engineering challenges to scale these systems for commericial use and make them competitive with traditional computers. This physics-based computing framework has a long history, and is gaining recent attention as an alternative computing paradigm. The design of Ising machines focuses on three key engineering challenges: 1) fast, small, and low-power oscillators, 2) a scalable, and expressive coupling architecture and 3) a generalizable algorithmic framework. Despite the challenges, there are many advantages to such a system. A physics-based processor can be significantly more energy efficient by leveraging the natural system dynamics as a computational resource. This has lead to extensive work developing a wide variety of analog, digital, and mixed-signal \cite{b5} implementations of coupled oscillators systems to address combinatorial optimization problems \cite{b6} and more general machine learning problems \cite{b7,b8,b9,b10,b11}.

\section{Methodology}

\begin{figure}[h]
    \centering
    \includegraphics[scale=0.5]{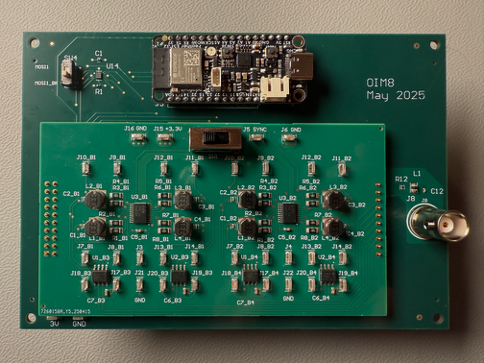}
    \raisebox{3mm}{\includegraphics[scale=0.3]{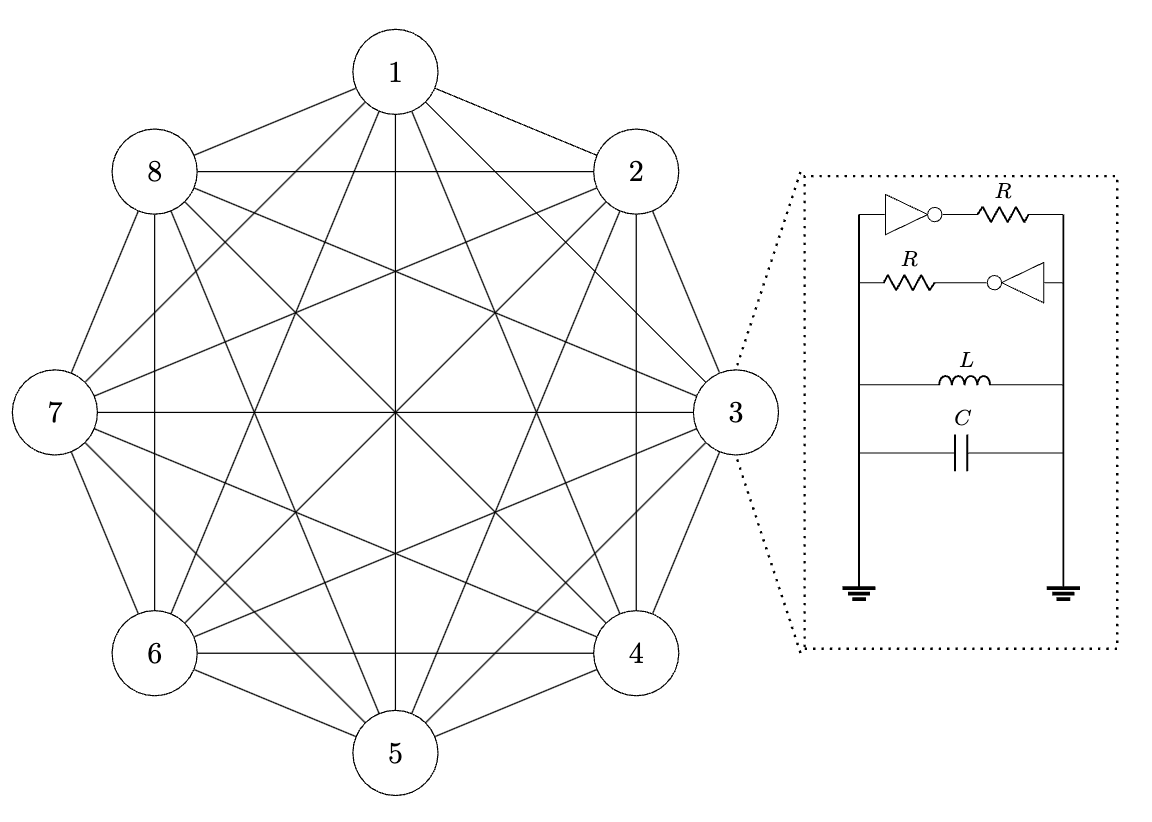}}
    \caption{The complete assembly of our circuit. It encodes the problem matrix using a network of inductor-capacitor (LC) oscillators that are fully-connected with programmable resistors.}
\end{figure}

This work contributes a mixed-signal PCB implementing the oscillator-based computing framework proposed in \cite{b12}. We further provide hardware details and a theoretical outlook that extends the proposed future research from the original work \cite{b12}. Fig. 1 shows our Oscillator Ising Machine (OIM) designed as an accelerator to minimize $\eqref{ising}$. The circuit has an inherent graph structure consisting of the top oscillator board that contains the spins $\mathbf{s}$ (nodes), and the bottom coupling board that encodes the problem matrix $\mathbf{J}$ (edges). The annealing control input is provided through the BNC connector using an external waveform generator. The ESP32 microcontroller handles digital control logic and reads measurements from the analog components. The circuit contains many testpoints and serves as an experimental platform to research oscillator-based computing and further explore energy-based machine learning frameworks. 

We focus on an electronic implementation based on current state-of-the-art research, as advanced design and manufacturing processes are well-established and competitive. Table 1 compares our work to the current state-of-the-art. While CMOS implementations on modern process nodes contain more spins and require less power consumption, they are very expensive and require long development timelines. Our circuit allows higher coupling precision than \cite{b11} using the same number of spin variables. While tunable resistive coupling was relatively straightforward using digital potentiometers, other approaches like capacitive coupling would be more power efficient in theory. All these devices make tradeoffs between the three key challenges outlined above.

\renewcommand{\arraystretch}{1}
\begin{table}[htbp]
\begin{center}
\caption{Comparison of existing electronic Ising machines}
\hspace*{-1cm}
\begin{tabular}{|p{0.4cm}|p{0.7cm}|p{1cm}|p{1cm}|p{1cm}|p{0.75cm}|p{1cm}|p{0.8cm}|}
\hline
\textbf{} & \textbf{\textit{Type}} & \textbf{\textit{Spins}} & \textbf{\textit{Coupling Graph}} & \textbf{\textit{Coupling Type}} & \textbf{\textit{Coupling Resolution}}  & \textbf{\textit{Frequency}} & \textbf{\textit{Power}} \\
\hline
Ours & PCB & 8 & Full & Resistive & 8-bit & 1 MHz & 200mW \\
\hline
\cite{b11} & PCB & 8 & Full & Capacitive & 2-bit signed & N/A & N/A \\
\hline
\cite{b26} & CMOS 65nm & 30 & Full & Capacitive & Fixed & 45 kHz & 2 mW \\
\hline
\cite{b28} & CMOS 65nm & 48 & Full & Gate & 4-bit & ~28.5 MHz & 16-105 mW \\
\hline
\cite{b12} & PCB & 240 & Chimera & Resisitive & 8 bits & 1 MHz & 5 W\\ 
\hline
\cite{b29} & CMOS 65mn & 512 & Full & Digital & 4-bit signed & 320 MHz & 649 mW \\
\hline
\cite{b35} & CMOS 28mn & 512 & Full & Digital & 4-bit & 1 MHz & 12 mW \\
\hline
\cite{b27} & CMOS 65nm & 560 & Hex & Inverter & Fixed & 118 MHz & 23 mW \\
\hline
\cite{b36} & CMOS 65mn & 600 & Full & Capacitive & Fixed & 45 kHz & 25 mW \\
\hline
\cite{b20} & CMOS 28nm & 1440 & Config & Current & 3-bit signed & 100 MHz  & 460 mW \\
\hline
\cite{b19} & CMOS 65nm & 1968 & King & Gate & 1-bit signed & 1 GHz & 42 mW \\
\hline
\cite{b31} & CMOS 65nm & 20k & 2D-Lattice & Digital  & 2-bit & 100 MHz & 57 mW \\
\hline
\cite{b31} & CMOS 40nm & 2x30k & King & Digital & 3-bit & 100 MHz & N/A \\
\hline
\cite{b33} & CMOS 40nm & 9x16k & Local & Digital & 5-bit & 100 MHz & N/A \\
\hline
\cite{b34} & CMOS 40nm & 9x9x16k & Local & Digital & 5-bit & 100 MHz & N/A \\
\hline
\end{tabular}
\end{center}
\end{table}

\subsection{Theory}

We begin by summarizing the operating principle of our circuit that descibes how a network of coupled nonlinear oscillators is used to minimize $\eqref{ising}$. Standard voltage analysis at the output of each oscillator node gives 

\begin{equation}
    K_c\mathbf{L}\mathbf{v} = \mathbf{i}\label{kcl}
\end{equation}

where $\mathbf{L} = \mathbf{D} - \mathbf{J} \in \mathbb{R}^{N \times N}$ is the weighted graph Laplacian, and $\mathbf{v} \in \mathbb{R}^{N}$ and $\mathbf{i} \in \mathbb{R}^{N}$ represent the voltage and current across the oscillators, respectively. The $K_c \in \mathbb{R}$ is a physical loading term that represents the global coupling bias. The input coupling weights are encoded as conductances $J_{ij}=\frac{1}{K_cR_{ij}} \in \mathbb{R}$ so that higher coupling values correspond to less resistance. When $J_{ij}$ is 0, oscillators $i$ and $j$ are disconnected. The current injection at each node induces a delay based on the characteristic phase response of the oscillator. The full derivation and analysis of the phase response model is out of scope for this work. As shown in \cite{b12}, the simplified phase response dynamics $\mathbf{\phi}(t) \in \mathbb{R}^{N}$ are described using the Kuramoto model

\begin{equation}
\frac{d\phi_i}{dt} = -K_c\sum_{j, j \ne i}^{N} J_{ij}\sin(\phi_i - \phi_j)\label{kura}
\end{equation}

The time dependence of $\phi$ is omitted for clarity. This has an associated Lyapunov function that describes the total "energy" 

\begin{equation}
    E(t) = -K_c\sum_{i,j, i \ne j}^{N} J_{ij}\cos(\phi_i - \phi_j)\label{lya}
\end{equation}

where $\frac{dE}{dt}\le 0$ and is lower-bounded such that the energy is minimized. If the phase of each oscillator settles to a binary value of $0$ or $\pi$, corresponding to $s$ equal to 1 or -1, then the dynamics of $\eqref{lya}$ minimize $\eqref{ising}$. Therefore, an external synchronization input (SYNC) is required to induce sub-harmoic injection locking (SHIL) that enforces these bi-stable phase configurations. By applying a sufficient amplitude waveform at twice the fundamental oscillator frequency, the phase dynamics should settle to either 0 or $\pi$ relative phase differences. In practice, the SYNC requires tuning and depends on perturbation characteristics predicted by the phase response model. The bifurcation and stability of oscillator networks has been well-studied in the context of simple theoretical Kuramoto models, but the analysis has rarely been extended to practical applications of combinatorial optimization. Under the influence of adjacent oscillators and an external SYNC input, the Kuramoto model representing the phase dynamics of the circuit is given by 

\begin{equation}
    \frac{d\phi_i}{dt} = -K_c\sum_{j, j \ne i}^{N} J_{ij}\sin(\phi_i - \phi_j) - K_s\sin(2\phi_i)\label{synckura}
\end{equation}

where $K_s \in \mathbb{R}$ is the effective amplitude of the SYNC input. This model also remarkably has a Lyapunov function given by

\begin{equation}
    E(t) = -K_c\sum_{i,j, i \ne j}^{} J_{ij}\cos(\phi_i - \phi_j) - K_s\sum_{i}^{N}\cos(2\phi_i)\label{synclya}
\end{equation}

that, under the binarizing effects of SHIL, reduces to

\begin{equation}
    E(t) \approx -K_c\sum_{i,j, i \ne j}^{} J_{ij}s_i s_j - N K_s \label{lyaising}
\end{equation}

which equals $\eqref{ising}$ with a constant offset so proper minimization is still possible. The coupling strength $K_{c}$ and the effective SYNC amplitude $K_{s}$ are the control inputs used for annealing. The $K_{s}$ can be controlled easily using an external waveform generator but tuning $K_{c}$ requires adjusting the global resistive bias in the network. For simplicity, our circuit does not allow $K_{c}$ to be adjusted. This restricts the possible annealing schedule, but for only 8 oscillators, our simulations indicate it shouldn't prevent them from achieving the best configuration. 

Overall, an oscillator-based architecture is a promising computation framework due to its inherent energy gradient descent dynamics. The Lyapunov function for the Kuramoto equations means that the system will settle to local minima of $\eqref{lyaising}$. However, as the Ising model $\eqref{ising}$ defines a discrete minimization problem, it may not have the same local minima as the continuous Lyapunov function. A fascinating future research direction is adapting an optimal control framework \cite{b14} for the annealing input based on minimum power dissipation \cite{b37} \cite{b17}. This control policy could optimize the overlap between minimization of the Lyapunov function and the Ising model. It has been shown that the oscillator system used in this work actually performs Lagrange Multiplier optimization \cite{b16} that minimizes the dissipated power across the resistors. This points to a deep connection with optimal control theory. Furthermore, Ising machines can be used for fast generation of Boltzmann statistics \cite{b18} and other training algorithms \cite{b13} fundamental to machine learning. Regardless of the theoretical framework, our circuit creates a constrained environment based on the input problem and utilizes the natural convergence of the phase dynamics to "search" the exponentially large space of solutions. This paradigm is fundamentally different than traditional algorithms run on classical digital computers. We now outline the core circuit components used to implement this theory in hardware.

\subsection{Oscillator Board}

\begin{figure}[h]
    \centering
    \includegraphics[scale=0.05]{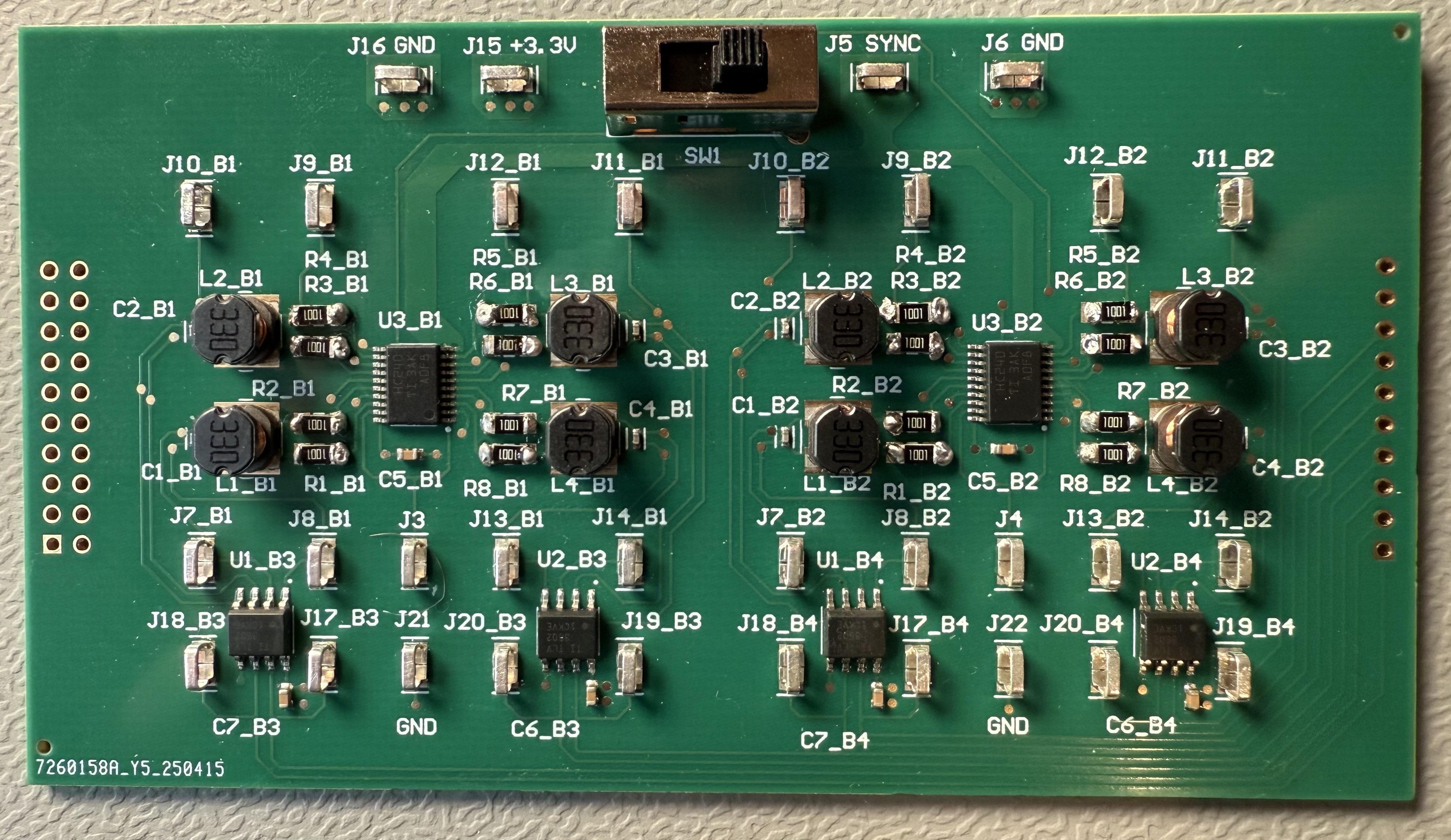}
    \caption{The oscillator board. The switch is used to toggle power to the oscillators.}
\end{figure}

Fig. 2 shows the standalone oscillator board that contains 8 independent LC tank oscillators to implement physical analogs of the logical spin variables. We previously considered ring oscillators as they require a smaller footprint but have less desirable SHIL characteristics \cite{b12}. Each oscillator is tuned to operate at a nominal 1MHz frequency and 3.3V amplitude, DC-offset by 1.65V due to the low-side grounded CMOS architecture. The left and right terminals of the oscillator are connected to a high-speed digital comparator configured as a zero-crossing detector. Two 0-degree coupled oscillators have approximately the same phase, so each oscillator's zero-crossing detector will read approximately identically. Therefore, without loss of generality, the binary spin variables are encoded in the relative phase of each oscillator.

\subsection{Coupling Board}

\begin{figure}[htbp]
    \centerline{\includegraphics[scale=0.06]{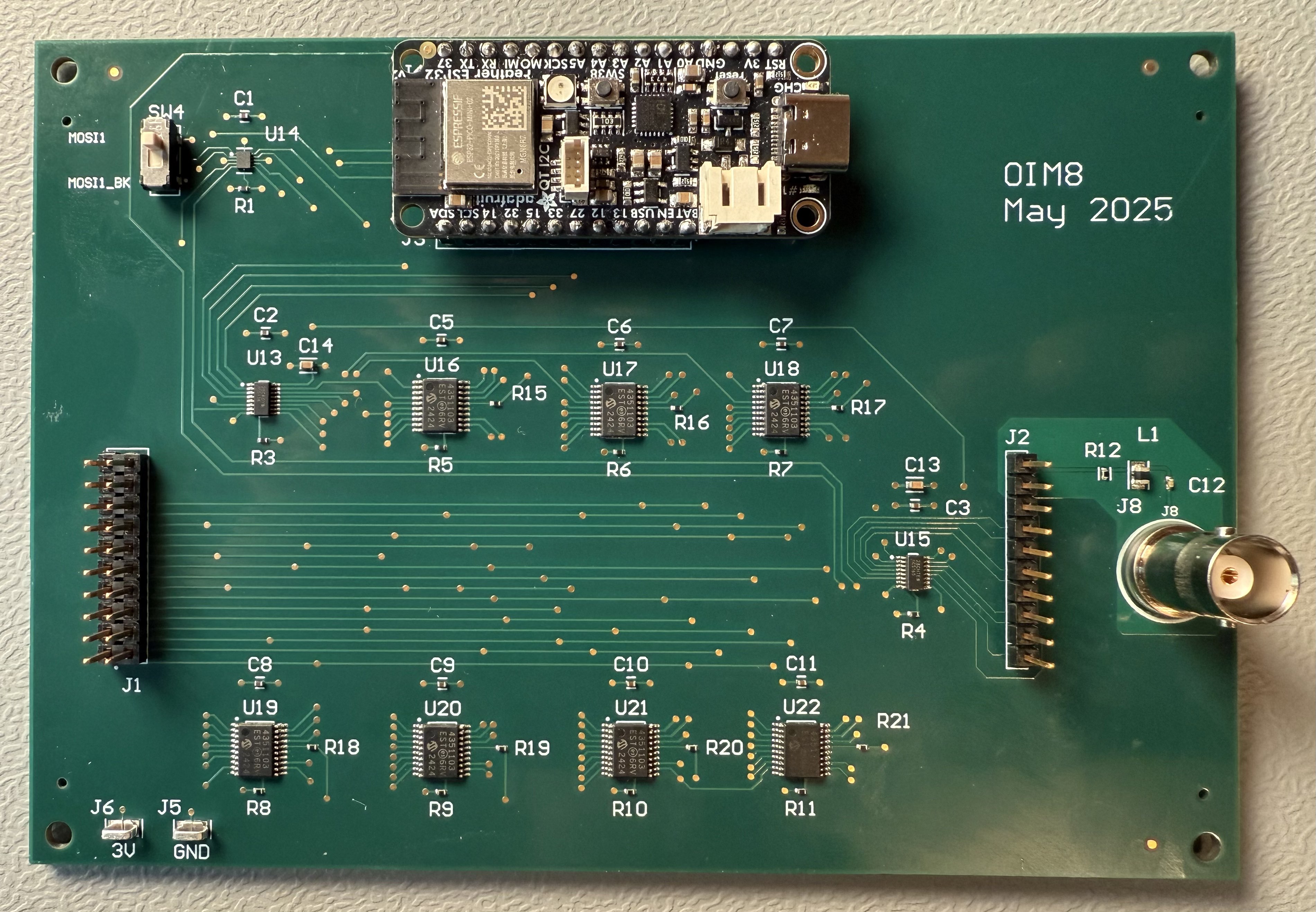}}
    \caption{The coupling board with a mounted ESP32 microcontroller.}
\end{figure}

Fig. 3 shows the coupling board along with the microcontroller. It uses digital potentiometers to fully-connect all 8 oscillators at opposite terminals. This pattern encourages anti-phase synchronization which allows proper minimization of $\eqref{ising}$. The oscillator board becomes connected when placed on the header pins of the coupling board. 

It's important to ensure a linear relation between the input problem matrix $\mathbf{J}$ and the set resistance of the network $\mathbf{R}$. Given the stringent requirement for linearity, we opted to use the MPC4351 from Microchip Semiconductors. This is an 8-bit quad-channel digital potentiometer, requiring 7 IC chips to make all 28 coupling connections. The potentiometer provides 257 resistive steps plus disconnection. The SPI protocol used by the digital potentiometer offers easy expandability as only one additional chip select pin is required for every additional digital potentiometer. To further reduce the GPIOs required from the microcontroller, a 3-8 decoder is used. As shown in Fig. 4, this allows communication with up to eight SPI devices using only seven GPIO pins from the ESP32 microcontroller.

\begin{figure}[h]
    \centering
    \includegraphics[scale=0.3]{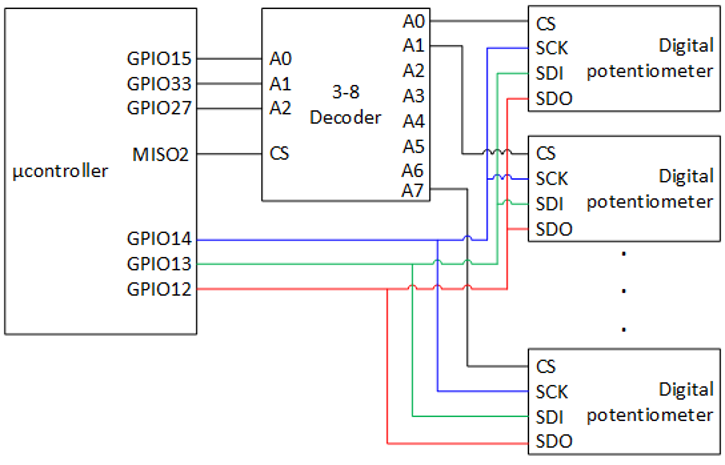}
    \caption{Block diagram connections between the microcontroller, the decoder, and the digital potentiometers.}
\end{figure}

We had initially decided against using digital potentiometers because of their nonlinear thermal behavior. Fig. 5 shows the two designs we originally considered, both based on a seemingly simpler R/2R resistor ladder with an analog switch network. The first R/2R parallel design connects a set of 4 analog switches to 4 resistors. This topology would allow for 16 levels of coupling, but results in nonlinear incremental resistance steps. Hence a second R/2R series design was considered, which has linear resistance steps, but was disregarded as it did not provide sufficient precision for coupling. Both of these approaches would have required significantly more area and complicated routing, all while mimicking a digital potentiometer. Another contributing factor to the decision to use the digital potentiometers was the board capacitances and inductances. Analysis of the analog switch footprint and layout showed a significant capacitance leading to a frequency bandwidth limitation much lower than the oscillator operating frequency. The track/trace and pad capacitances would be too high for the designed oscillator frequency. Therefore, although the digital potentiometers have nonlinear behavior, they were choosen in favor of the simplier implementation. A standard error analysis of the digital potentiometers showed the nonlinear effects were small, and could be neglected for our circuit.

\begin{figure}[h]
    \centering
    \includegraphics[scale=0.25]{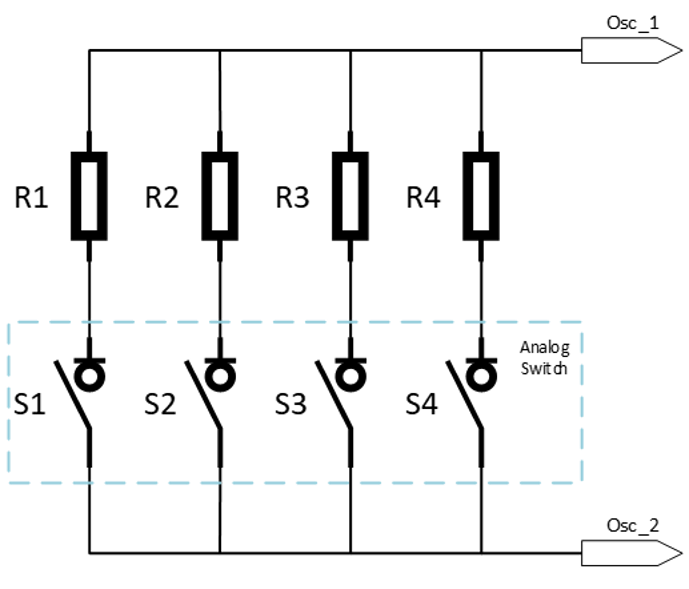}
    \includegraphics[scale=0.25]{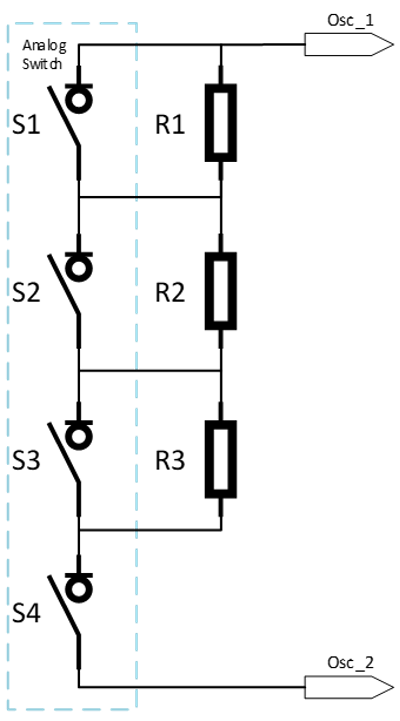}
    \caption{The two R/2R resistor ladder designs considered for implementing a single tunable resistor (e.g., $R_{12}$). Neither of the designs were used in the circuit.}
\end{figure}

\subsection{Synchronization}

As outlined above, the phase dynamics, which ideally minimize the Lyapunov function $\eqref{lyaising}$, do not necessarily minimize the Ising model (1) representing the input problem. This is because the Lyapunov function uses continuous phases values, whereas the Ising model uses discrete variables. Therefore, the external SYNC enforces binary phase configurations, but crucially does not change their relative positions and keeps the energy landscape over all phase configurations unchanged. The BNC connector fixed to the coupling board is used to supply the SYNC input via an external function generator to allow flexible annealing schedules. A high-pass filter built into the coupling board removes undesirable noise from the BNC input. The filter introduces a small phase delay, however since all oscillators undergo the same SYNC, it should not affect performance. We found the required SYNC amplitude to be roughly 400mV, although more experimental calibration is needed.

\subsection{Programming}

The Arduino IDE environment was used to program the ESP32. An overview of the pseudocode used to operate the circuit is shown below. We use several measurements (e.g., 10) to estimate the phase configuration once the oscillators settle. A comprehensive benchmark using a variety of graph structures and quantized weights is left for future work. 

\lstset{
  basicstyle=\ttfamily\small,
  backgroundcolor=\color{gray!10},
  frame=single,
  keywordstyle=\color{blue},
  commentstyle=\color{green},
  stringstyle=\color{red},
}
\begin{lstlisting}[language=C++]
#include < SPI.h >
// Define Pins
// Define Addresses
void setup() {
// Initialize GPIO pins
hspi = new SPIClass(HSPI);
hspi->(...)
test();
}

void test() {
// -1 (disconnected), 0 (lowest), 256 (highest)
  int16_t R[28] = {
    127, 127, 127, 127, 127, 127, 127,
    127, 127, 127, 127, 127, 127,
    127, 127, 127, 127, 127,
    127, 127, 127, 127,
    127, 127, 127, 
    127, 127,
    127
  };

  // Set 28 couplings
  for (uint8_t addr=0; addr < 7; addr++) {
    setDigiPot(addr, &(R[4*addr]));
  }

  // Read 10 phase samples
  byte results[10];
  for (int i=0; i < 10; i++) {
    results[i] = readShiftRegister(hspi);
  }
}
\end{lstlisting}

\section{Evaluation}

\subsection{Simulation}

\begin{figure}[h]
    \centering
    \includegraphics[scale=0.11]{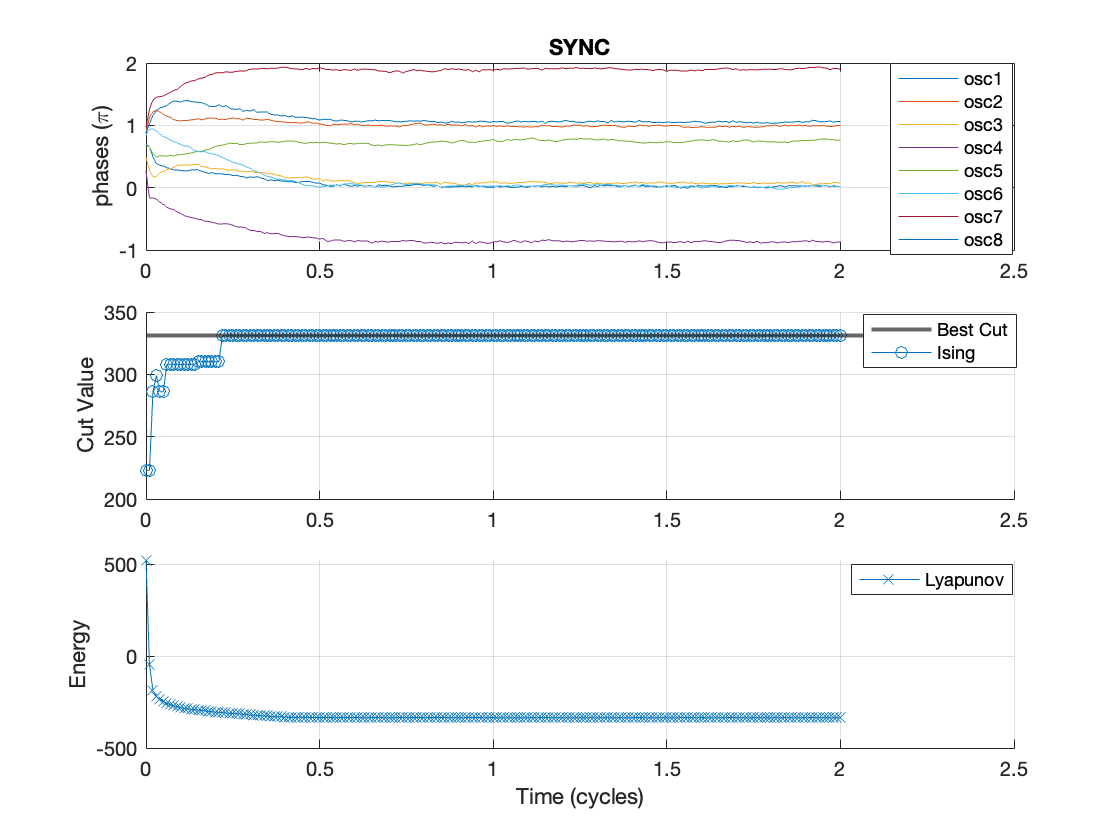}
    \includegraphics[scale=0.11]{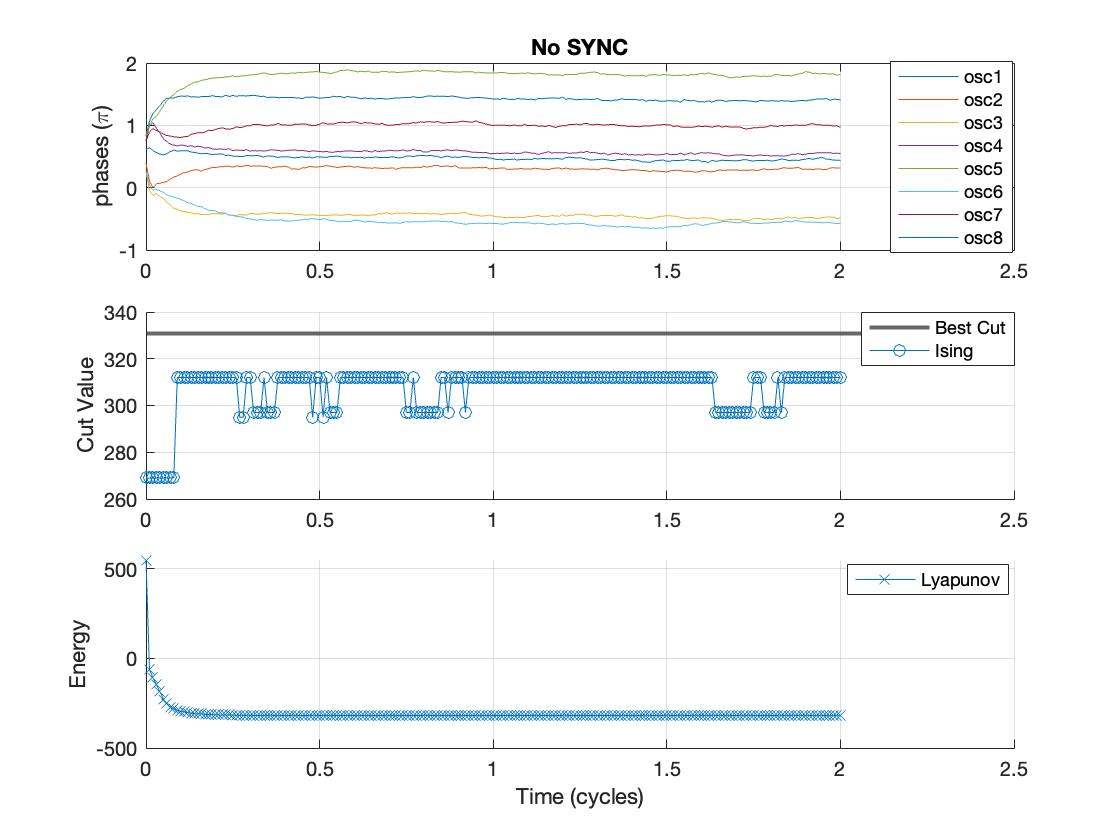}
    \caption{Comparison of phase dynamics solving a random Max-Cut instance with (left) and without (right) the external SYNC contribution. Note the top subplot, where SYNC causes the phases to settle around integer multiples of $\pi$, whereas the phases without SYNC settle to a spectrum of values.}
    \label{fig:yourlabel}
\end{figure}

The Kuramoto $\eqref{synckura}$ and the associated Lyapunov $\eqref{synclya}$ functions were simulated using the MATLAB $\texttt{sde}$ framework for stochastic differential equations. Some noise helps the oscillators settle to the ideal phase configuration, akin to how simulated annealing and similar methods utilize randomness to avoid local minima. Fig. 6 shows an example MaxCut graph problem, which sets $\mathbf{J}=-\mathbf{A}$ using the adjacency matrix $\mathbf{A}$ of the graph to be cut. As mentioned above, our circuit does not support tuning $K_c$ so its kept constant in this simulation. We find that the phases can still reach the best configurations reliably, indicating that a non-tunable physical $K_c$ isn't an issue for our circuit. However, a circuit with more spin variables would be restricted. Fig. 6 highlights the importance of the SYNC input because ultimately a binary solution is required. The optimal cut values were found using the MATLAB \texttt{qubo} interface, which provides the classical Tabu search algorithm. 

\begin{figure}[h]
    \centering
    \includegraphics[scale=0.4]{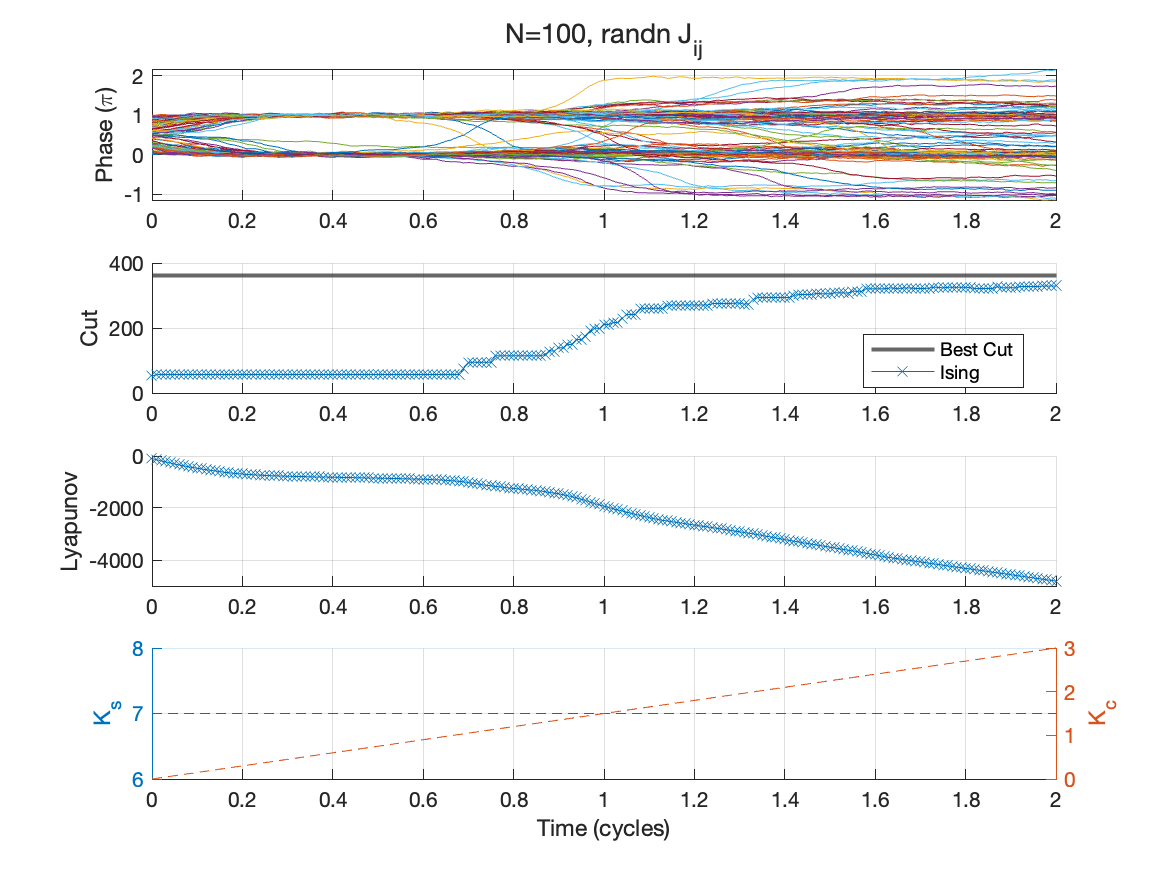}
    \caption{Simulation of 100 oscillators to solve a random MaxCut instance with standard normally distributed weights.}
\end{figure}

Fig. 7 depicts the phase dynamics for a larger 100-variable MaxCut problem instance. The phase of each oscillator settles near an integer multiple of $\pi$ due to the injection locking (subplot 1). The annealing schedule (subplot 4) ramps the coupling gain $K_{c}$ and keeps the synchronization gain $K_{s}$ fixed using empirically chosen values. A key future research direction is a principled annealing framework for Kuramoto networks applied to combinatorial optimization. The network almost reaches an optimal phase configuration corresponding to the optimal cut or groundstate (subplot 2). The Lyapunov function is gradually minimized (subplot 3) even when the corresponding cut value remains constant. This is evident around 0.8 cycles. Note these oscillators generally don't undergo many phase transitions (bitflips), yet reliably move the system towards the optimal configuration. Therefore, for oscillator-based computing systems, comparing the commonly used bitflips/Watt metric may not be a fair benchmark against other Ising machine technologies.

While the MATLAB simulations closely follow the theoretic framework, they do not necessarily describe the circuit-level dynamics. LTspice was used to simulate the noiseless LC oscillators. Exact vendor-specific models for our circuit components were not available which limited the utility of the analysis. For example, the inverters were simulated by taking the propagation and rise time of the IC from the datasheet and importing these values into LTspice. This approach provides a more realistic simulation to characterize the oscillator quality. Fig. 8 depicts how two oscillators are coupled, and Fig. 9 shows their reliable independent startup behavior and coupling.

\begin{figure}[h]
    \centering
    \includegraphics[scale=0.25]{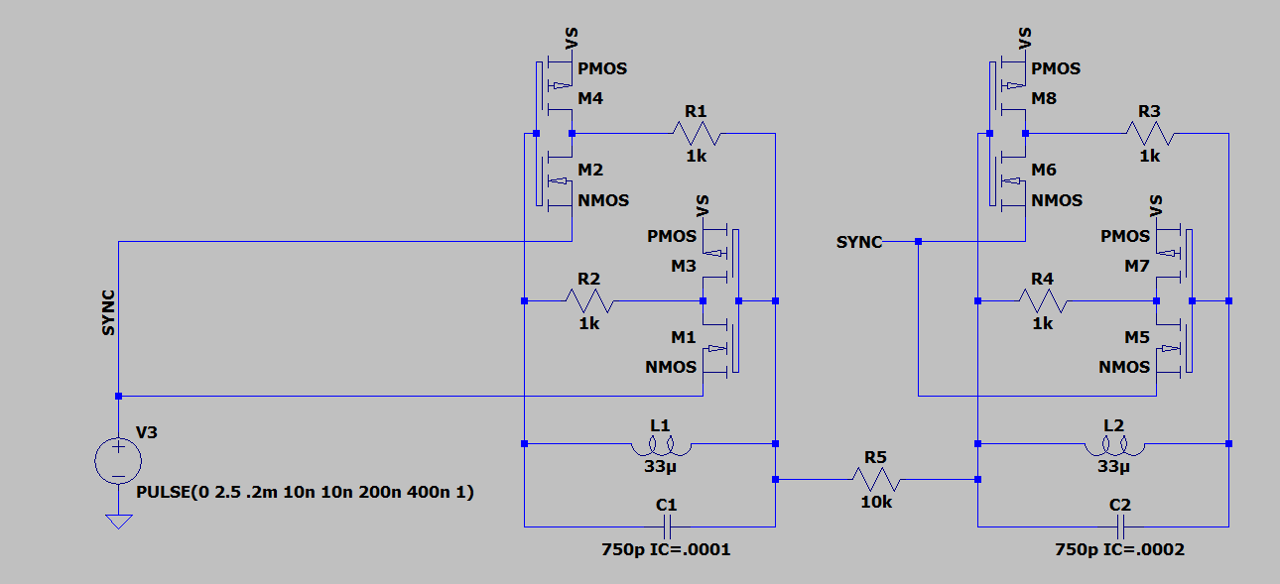}
    \caption{LTspice model depicting two coupled LC oscillators with the SYNC input at the ground of the CMOS components.}
\end{figure}

\begin{figure}[h]
    \centering
    \includegraphics[scale=0.3]{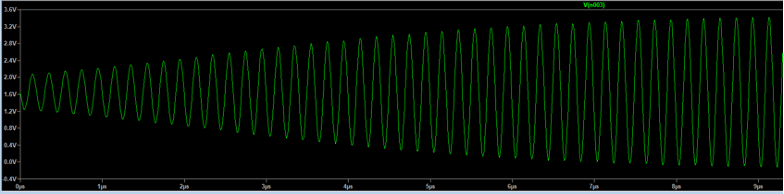}
    \includegraphics[scale=0.4]{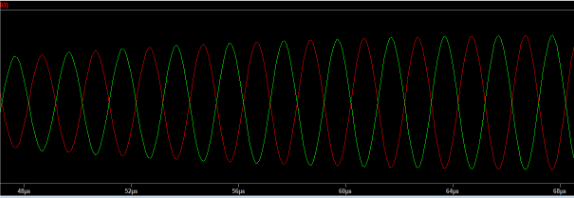}
    \caption{LTspice simulations comparing the startup of a single independent oscillator (top) and the behavior of two connected oscillators (bottom).}
\end{figure}

\subsection{Experiments}

Prior to designing the PCB, a breadboard prototype with 4 LC oscillators was used to verify and demonstrate the fundamental operating principle. The experimental setup is shown in Fig. 10. The oscillators use the dual in-line package version of the CMOS inverter component on the oscillator circuit board, are tuned to the same 1MHz frequency, and follow the same schematic in Fig. 8 (without the SYNC input). Not shown is a small decoupling capacitor placed on the breadboard power rails as a basic filter.

\begin{figure}[h]
    \centering
    \includegraphics[scale=0.5]{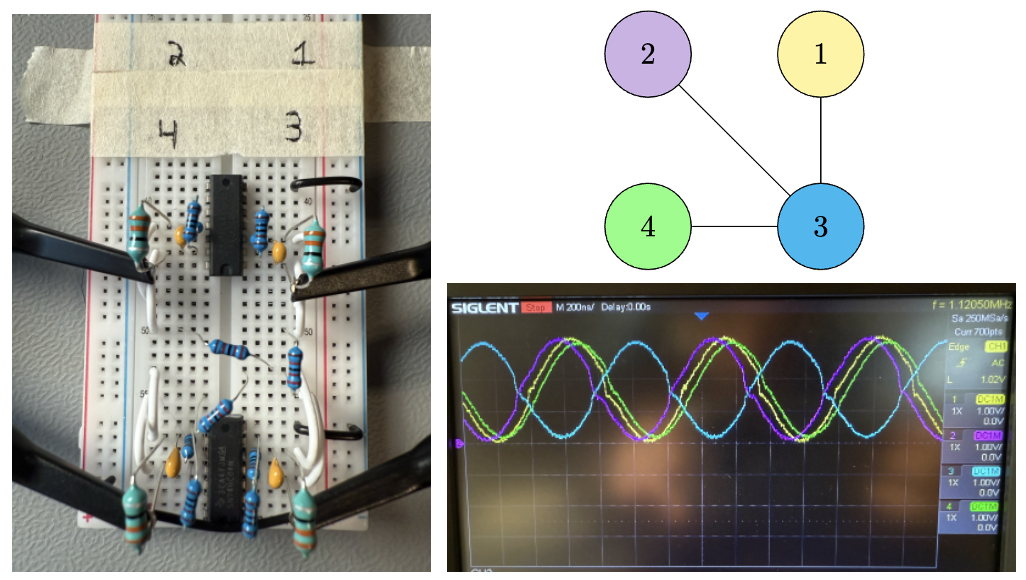}
    \caption{Breadboard demonstration of 4 LC oscillators solving a MaxCut problem. The colors match each node to the corresponding waveform on the oscilloscope. No external SYNC input is required to ensure bistable phases.}
\end{figure}

In this case, discrete 10K ohm resistors connect spin $s_3$ (blue) to spins $s_1$ (yellow), $s_2$ (purple), and $s_4$ (green) which themselves share no connections. Identical probes are attached at the same terminal of the inductor at each oscillator. The oscilloscope shows the set of spins $\{s_1, s_2, s_4\}$ are mis-aligned with $\{s_3\}$, which indicates the global minimum solution of $\mathbf{s}=[1, 1, -1, 1]$ or equivalently $\mathbf{s}=[-1, -1, 1, -1]$. In the context of the MaxCut problem, this corresponds to "cutting all the resistors" which forms the optimal partition for this graph since all edges contribute a positive weight.

To characterize the quality of the oscillator board, we measured their waveforms without any coupling. Fig. 11 depicts two disconnected oscillators (yellow, green) with their respective window comparator outputs (blue, purple). When the oscillators are disconnected, they are expected to be fully aligned which is evident, despite a small amount of phase delay. There is a small amount of distortion in the waveform. This is likely caused from human error in soldering additional resistors atop the oscillator board, and component and manufacturing variability. The fundamental frequency still is very close to the intended 1MHz.

\begin{figure}[htbp]
    \centerline{\includegraphics[scale=0.06]{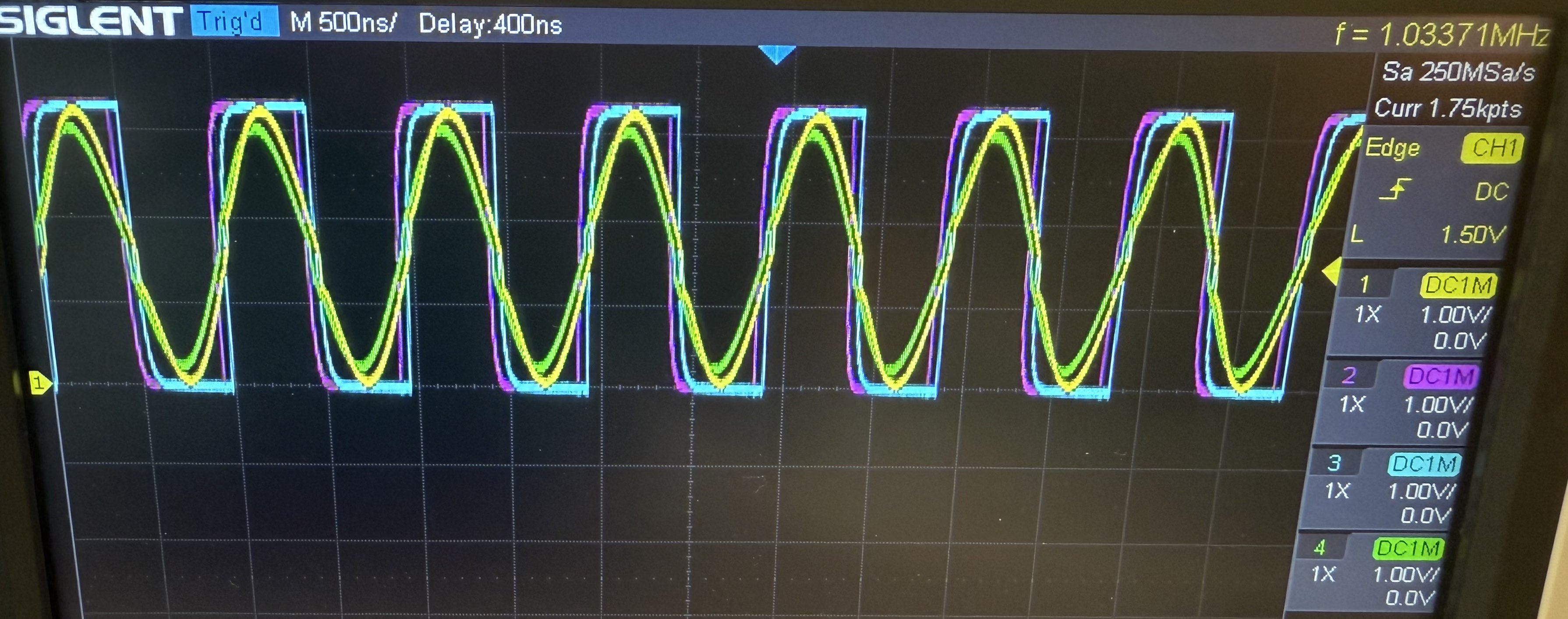}}
    \caption{Oscilloscope output from 2 of the 8 total waveforms and their window comparator outputs from the standalone oscillator board. All 8 waveforms and comparators outputs were measured on the oscilloscope, these two were choosen arbitrary to show. All oscillators and comparators had similar distortion.}
\end{figure}

\section{Discussion}

We designed the circuit with separate boards in anticipation of possible manufacturing delays or defects. Due to high unexpected parasitic capacitances, the oscillators did not always startup. To fix this, we hand-soldered additional resistors in parallel on top of the oscillator board to lower the internal resistance. A detailed analysis of the phase response characterization for the LC oscillator was outside the scope of this report.

Providing empirical results to demonstrate the energy-efficiency of the circuit against traditional computers is outside the scope of this report and will be the focus of future work. Using a similar architecture, it's projected that around 200 oscillators are required to reach the cross-over point to be competitive with a traditional computer for minimizing the Ising model (1) \cite{b15}.

The coupling board presented here does not support diagonal terms in $\mathbf{J}$, which correspond to independent resistive biases on each oscillator. This was intentional to simplify the design and avoid more complicated layout and routing. It also does not directly support signed coefficients, although this could be addressed in software with proper scaling.  

In principle, the global coupling strength $K_c$ of the network could be controlled but is kept fixed in our circuit for simplicity. Formulating the SYNC input as an optimal control problem is an interesting future research direction.

\section{Contributions}

Bowring proposed the general circuit designs, ran the breadboard experiments, developed MATLAB and LTspice simulations, wrote the manuscript, and contributed to the C++ firmware. Tiffany developed the coupling board schematic, wrote the initial C++ firmware, and managed the manufacturing and shipping details. Anderson contributed to the board layouts and routing.

\end{document}